\documentclass[12pt]{article}

\ifx\pdfoutput\undefined
\usepackage[dvips,bookmarks]{hyperref}
\else
\usepackage{hyperref}
\fi
\hypersetup{colorlinks=false,bookmarksopen,bookmarksnumbered,citecolor=blue,
   pdfstartview=FitH}

\usepackage{latexsym}
\usepackage{amssymb,amsfonts,amsmath}
\usepackage{graphicx} 
\usepackage{indentfirst}
\usepackage{bbm}
\parskip=\medskipamount

\newcommand {\cD}{{\cal D}}
\newcommand {\cE}{{\cal E}}

\newcommand {\cH}{{\cal H}}

\newcommand {\cJ}{{\cal J}}

\newcommand {\cN}{{\cal N}}
\newcommand {\cO}{{\cal O}}

\newcommand {\cZ}{{\cal Z}}
\def\a{\alpha}
\def\b{\beta}

\def\d{\delta}

\def\g{\gamma}

\def\l{\lambda}

\def\q{\theta}

\def\s{\sigma}

\def\D{\Delta}

\def\O{\Omega}

\def \bi{\bibitem}

\def\ri{{\rm i}}

\newcommand{\ad}{{\dot{\alpha}}}                           %new
\newcommand{\bd}{{\dot{\beta}}}                            %new
\newcommand{\ve}{\varepsilon}                            %new
\newcommand{\cDB}{{\bar\cD}}                            %new

\newcommand{\pa}{\partial}                           %new
\newcommand{\hf}{\frac12}
\newcommand{\be}{\begin{equation}}
\newcommand{\ee}{\end{equation}}
\newcommand{\bea}{\begin{eqnarray}}
\newcommand{\eea}{\end{eqnarray}}
\newcommand{\non}{\nonumber}
\newcommand{\ba}{\begin{array}}
\newcommand{\ea}{\end{array}}

\newcommand{\1}{\underline{1}}
\newcommand{\2}{\underline{2}}

\newcommand{\dsC}{{\mathbb C}}

\newcommand{\bm}[1]{\mbox{\boldmath$#1$}}

\def\double #1{#1{\hbox{\kern-2pt $#1$}}}

\newcommand{\gd}{{\dot\g}}
\newcommand{\dd}{{\dot\d}}

\newcommand{\sba}{{\bar{\s}}}

\newcommand{\rd}{\mathrm d}
\begin{document}

\begin{titlepage}
\begin{flushright}
%January 2012
~\\
\end{flushright}
\vspace{5mm}

\begin{center}
{\Large \bf
Nonlinear  self-duality in $\bm{\cN = 2}$ supergravity}
\\ 
\end{center}

\begin{center}

{\bf Sergei M. Kuzenko }

\footnotesize{
{\it School of Physics M013, The University of Western Australia\\
35 Stirling Highway, Crawley W.A. 6009, Australia}}  ~\\
\texttt{sergei.kuzenko@uwa.edu.au}\\
\vspace{2mm}

\end{center}
\vspace{5mm}

\begin{abstract}
\baselineskip=14pt
For nonlinear models of an Abelian vector supermultiplet coupled to $\cN=2$ supergravity in four dimensions, 
we formulate the self-duality equation which expresses 
invariance under U(1) duality rotations. 
In the flat space limit, this equation reduces to the $\cN=2$ self-duality equation proposed  in hep-th/0001068.
We also give an example of a self-dual locally supersymmetric model containing
a higher-derivative extension of  the Born-Infeld action 
at the component level.
\end{abstract}
%\vspace{1cm}

\vfill
\end{titlepage}

\newpage
\renewcommand{\thefootnote}{\arabic{footnote}}
\setcounter{footnote}{0}

%\tableofcontents

%\pagestyle{plain}
%\pagenumbering{arabic}

% Equation numbers
%\numberwithin{equation}{section}

%\allowdisplaybreaks

%===BEGIN DOCUMENT=============================================================
%\section{Introduction}
\setcounter{footnote}{0}

Given a model for nonlinear electrodynamics described by a Lagrangian $L(F_{ab})$, 
its invariance under U(1) duality rotations is known to be equivalent to the requirement that 
the Lagrangian should obey the self-duality equation 
\be
G^{ab}\, \tilde{G}_{ab} +F^{ab} \, \tilde{F}_{ab} = 0~,
\label{GZ}
\ee
where
\be
\tilde{G}_{ab} (F):=
\hf \, \ve_{abcd}\, G^{cd}(F) =
2 \, \frac{\pa L(F)}{\pa F^{ab}}~,\qquad
G(F) =  \tilde{F} + \cO(F^3)~.
\label{tilde-g}
\ee
This equation was originally derived by Gibbons and Rasheed in 1995 \cite{GR1}.
Two years later, it was re-derived by  Gaillard and Zumino  \cite{GZ2} building on 
on their 1981 work  \cite{GZ1}.
Such self-dual theories possess  interesting properties \cite{GZ2,GZ3}
reviewed in \cite{KT2} (see also \cite{AFZ} for a more recent review). 
In particular, the action functional is automatically invariant under Legendre transformation.
The self-duality equation (\ref{GZ}) can be re-formulated to be suitable for theories with higher derivatives 
 \cite{KT2}.

The concept of self-dual nonlinear electrodynamics was generalized  to the cases
of $\cN=1$ and $\cN=2$ rigid supersymmetric theories in \cite{KT1}.
This generalization has turned out to be very useful, since the families of actions 
obtained include all the known models 
for partial breaking of supersymmetry based 
on the use of a vector Goldstone multiplet.
In particular, the $\cN=1$ supersymmetric Born-Infeld
action \cite{CF}, which  is a Goldstone multiplet 
action for partial supersymmetry 
breakdown $\cN=2 \to \cN=1$ \cite{BG,RT} is, at the same time,   a solution 
to the $\cN=1$  self-duality equation 
\cite{KT2,KT1}.  Furthermore, the model for partial 
breaking of supersymmetry $\cN=4 \to \cN=2$  \cite{BIK}, which nowadays is identified
with the $\cN=2$ supersymmetric Born-Infeld action, 
was first constructed in \cite{KT2}
as a unique solution to  the $\cN=2$ self-duality equation 
possessing a nonlinearly realized central charge 
symmetry. 

Models for self-dual nonlinear $\cN=1$ supersymmetric electrodynamics \cite{KT2,KT1}  were generalized  to 
supergravity in \cite{KMcC}.  For several years, however, 
an extension to the case of $\cN=2$ supergravity 
was not feasible to achieve,  due to non--existence of a useful superspace formulation for $\cN=2$
supergravity-matter systems (although bits and pieces of curved superspace constructions
had been known for quite a while).
 Such a formulation has recently been developed  \cite{KLRT-M1,KLRT-M2}. 
Using the results of these and related  works \cite{KT-M,BK}, in this note we present a general setting 
for duality invariant theories of an Abelain $\cN=2$ vector multiplet coupled to $\cN=2$ supergravity. 
Throughout this paper,  we will use  the  superspace formulation for $\cN=2$ conformal supergravity
developed in \cite{KLRT-M1} and based on the curved superspace geometry 
introduced by Grimm \cite{Grimm};  its salient points are summarized in the Appendix.
Our results can naturally be extended 
to more  general superspace formulations 
for $\cN=2$ conformal supergravity, with  larger structure groups of the curved superspace, 
 which were developed by Howe \cite{Howe}
and  Butter \cite{Butter}.

The Abelian vector multiplet  coupled to $\cN=2$ conformal supergravity can be described 
by its covariantly chiral field strength $W$, 
\bea
\cDB_{\ad i} W= 0~, 
\label{3}
\eea
  subject to the Bianchi identity\footnote{Such a superfield 
is often called reduced chiral.} \cite{KLRT-M1,Howe}
\bea
\Big( \cD^{ij} + 4S^{ij}\Big) W&=&
\Big(\bar \cD^{ij} +  4\bar{S}^{ij}\Big)\bar{W} ~,
\label{n=2bi-i}
\eea
where  $\cD^{ij}:= \cD^{\a(i}\cD_\a^{j)}$ and 
$\bar \cD^{ij} :=  \cDB_\ad{}^{(i}\cDB^{j) \ad}$; 
$S^{ij} $ and its conjugate ${\bar S}_{ij} =\ve_{ik} \ve_{jl} \bar S^{kl}$  
are special dimension-1 components of the torsion, see the Appendix.
In the flat superspace limit these relations reduce to those given in \cite{GSW}.

There are several ways to realize $W$ as a gauge invariant field strength.
In this paper, we will solely use
a curved-superspace extension 
of Mezincescu's prepotential \cite{Mezincescu} (see also \cite{HST}),  $V_{ij}=V_{ji}$,
which is an unconstrained real SU(2) triplet, $(V_{ij})^* =\ve^{ik}\ve^{jl}V_{kl}$. 
The expression for $W$ in terms of $V_{ij}$ 
was shown in \cite{BK} to be
\begin{align}
W = 
\bar\Delta \Big({\cD}^{ij} + 4 S^{ij}\Big) V_{ij}~.
\label{Mez}
\end{align}
Here $\bar\Delta$ is the covariantly chiral projection operator \cite{Muller}
\bea
\bar{\D}
&=&\frac{1}{96} \Big((\cDB^{ij}+16\bar{S}^{ij})\cDB_{ij}
-(\cDB^{\ad\bd}-16\bar{Y}^{\ad\bd})\cDB_{\ad\bd} \Big)
\non\\
&=&\frac{1}{96} \Big(\cDB_{ij}(\cDB^{ij}+16\bar{S}^{ij})
-\cDB_{\ad\bd}(\cDB^{\ad\bd}-16\bar{Y}^{\ad\bd}) \Big)~,
\label{chiral-pr}
\eea
where $\cDB^{\ad\bd}:=\cDB^{(\ad}_k\cDB^{\bd)k}$; 
 $\bar{Y}^{\ad\bd}$ is a special 
dimension-1 component of the torsion, see the Appendix.
The  fundamental property  of $\bar \D$ is that $\bar{\D} U$ is covariantly chiral,
for any scalar and isoscalar superfield $U$,
that is ${\bar \cD}^{\ad}_i \bar{\D} U =0$. 
This operator relates an integral over the full superspace to that over its chiral subspace:
\bea
\int \rd^4 x \,{\rm d}^4\q \,{\rm d}^4{\bar \q}\,E\, U 
=  \int \rd^4 x \,{\rm d}^4\q \,
\,\cE\, \bar\D U 
~, \qquad E^{-1}= {\rm Ber}(E_A{}^M)~,
\label{chiral_action_rule}
\eea
with $\cE$ the chiral density, see \cite{KT-M} for a derivation.

Let $ S[W , {\bar W}]$ be an action functional describing the dynamics
of the $\cN=2$ vector multiplet. Suppose that $S[W , {\bar W}]$ can be unambiguously
defined as a functional of {\it unconstrained}
(anti) chiral superfields ${\bar W}$ and $W$.
Then, one can introduce a covariantly chiral superfield
 $M$ as
\be
{\rm i}\, M := 4\, \frac{\d }{\d W}\,
S[W , {\bar W}]
~,  \qquad \cDB_{\ad i} M= 0~, 
\label{n=2vd}
\ee
where the variational derivative $\d S/\d W$ is defined by 
\bea
\d S =  \int \rd^4 x \,{\rm d}^4\q \,\cE\, \d W \frac{\d S}{\d W}~+~{\rm c.c.}~.
\eea
In terms of $M$ and its conjugate $\bar M$,  the equation of motion for $V_{ij}$ is
\bea
\Big(\cD^{\a(i}\cD_\a^{j)}+4S^{ij}\Big) M&=&
\Big(\cDB_\ad{}^{(i}\cDB^{j) \ad}+4\bar{S}^{ij}\Big)\bar{M} ~.
\label{n=2em}
\eea
Here we have used the representation (\ref{Mez}).

Consider an infinitesimal  super-Weyl transformation of the covariant derivatives \cite{KLRT-M1}
given by
\begin{subequations}\label{11}
\bea
\d_{\s} \cD_\a^i&=&\hf\sba\cD_\a^i+(\cD^{\g i}\s)M_{\g\a}-(\cD_{\a k}\s)J^{ki}~, \\
\d_{\s} \cDB_{\ad i}&=&\hf\s\cDB_{\ad i}+(\cDB^{\gd}_{i}\sba)\bar{M}_{\gd\ad}
+(\cDB_{\ad}^{k}\sba)J_{ki}~, 
\label{super-Weyl1} 
\eea
\end{subequations}
where  the parameter $\s$ is an arbitrary covariantly chiral superfield, $\bar \cD^\ad_i \s =0$.
The Lorentz generators, $M_{\a\b}$ and $\bar M_{\ad \bd}$, and the SU(2) generators, $J_{ij}$, 
are defined in the Appendix. 
Under the transformation \eqref{11}, $W$ varies as \cite{KLRT-M1}
\be
\d_{\s} W = \s W~.
\label{Wsuper-Weyl}
\ee
This transformation law is induced by 
the following variation of Mezincescu's prepotential:
\bea
\d_\s V_{ij} = -(\s +\bar \s) V_{ij}~.
\eea 
We assume that the action $ S[W , {\bar W}]$ is super-Weyl invariant 
$ \d_\s  S[W , {\bar W}] =0$. Making use of \eqref{Wsuper-Weyl} and 
the super-Weyl transformation of the chiral density\footnote{The full superspace density, $E$, is inert
 under the super-Weyl transformations \cite{KLRT-M1}, $\d_\s E =0$. 
The left-had side of \eqref{chiral_action_rule} is super-Weyl invariant if $\d_\s U=0$. 
This implies  that $\d_\s (\bar \D U )= 2 \s \bar \D U$.} \cite{KLRT-M1},
$\d_\s \cE = -2 \s \cE$, 
we obtain the super-Weyl transformation of $M$:
\be
\d_{\s} M = \s M~.
\label{Msuper-Weyl}
\ee

Since the Bianchi identity (\ref{n=2bi-i}) and the equation of
motion (\ref{n=2em}) have the same functional form,
and also since the super-Weyl transformation laws \eqref{Wsuper-Weyl} 
and \eqref{Msuper-Weyl} are identical, 
one can consider infinitesimal U(1) duality transformations
\be
\d W = \l \, M~, \qquad 
\d M  = -\l \, W~,
\label{n=2dt}
\ee
with $\l$ a constant parameter.
In complete analogy with the rigid supersymmetric case \cite{KT2,KT1}, 
the theory with action $ S[W , {\bar W}] $
can be shown to be duality invariant if the following reality condition holds
\bea
{\rm Im} \int \rd^4 x \,{\rm d}^4\q \,\cE\,  \Big( W^2 + M^2 \Big) =0~.
\label{SDE}
 \eea
In the flat superspace limit, this   reduces to the $\cN=2$ self-duality equation \cite{KT2,KT1}.

Any solution $ S[W , {\bar W}] $ of the self-duality equation \eqref{SDE} describes a vector multiplet model 
in curved superspace which is invariant under U(1) duality rotations. Properties of such locally supersymmetric theories are analogous to those in flat superspace \cite{KT2,KT1}. 
The key observation is that 
the action itself is not duality invariant, but
\be
\delta\, \Big(S- \frac{ {\rm i}}{8}\int \rd^4 x \,{\rm d}^4\q \,\cE\,    W \,M
+\frac{{\rm i}}{ 8}\int \rd^4 x \,{\rm d}^4 \bar \q \, \bar \cE\, \bar W \bar M
\Big) ~=~0~.
\ee
The invariance of the latter functional under a finite U(1)
duality rotation by $  \pi / 2$,
is equivalent to the self-duality
of $S$ under Legendre transformation,
\be
S[ W, {\bar W} ]
- {{\rm i}\over 4} \int \rd^4 x \,{\rm d}^4\q \,\cE\, 
 W W_{\rm D}
+ {{\rm i}\over 4}\int \rd^4 x \,{\rm d}^4 \bar \q \, \bar \cE\, 
{\bar W} {\bar W}_{\rm D}
=S[W_{\rm D}, {\bar W}_{\rm D} ]~,
\ee
where $W_{{\rm D}}$ is
the dual chiral field strength,
\be
W_{\rm D} =  \bar \D ( \cD_{ij} +4S^{ij} ) \, V_{\rm D}{}^{ij}~,
\label{dfs}
\ee
with $V_{\rm D}{}^{ij}$ a real unconstrained prepotential.

Suppose that the action of our self-dual theory, $S[ W, {\bar W} ; g ]$,  
depends on a duality invariant parameter $g$.  Then, the functional $\pa S /\pa g$ is duality invariant. 
The proof of this result is analogous to 
the non-supersymmetric or rigid supersymmetric cases, see e.g. \cite{KT2}. 
Let us consider the supercurrent of the theory, 
\bea
\cJ  = \frac{\d S}{\d \cH}~,
\eea
where $\cH$ is the real scalar  prepotential describing the Weyl multiplet of $\cN=2$ supergravity, 
see \cite{KT} for more details. Since $\cH$ is duality-invariant, we conclude that the supercurrent of any 
self-dual theory is duality invariant.

We now give an example of a self-dual theory. It is described by the action 
\bea
S= \frac{1}{4}  \int \rd^4 x \,{\rm d}^4\q \,\cE\,  X 
+  \frac{1}{4}  \int \rd^4 x \,{\rm d}^4 \bar \q \, \bar \cE\,  \bar X ~, 
\label{21}
\eea
where the chiral superfield $X$ is a functional
of $W$ and $\bar W$ defined via the
constraint
\be
X = \frac{X}{\cZ^2} \, {\bar \D} \frac{{\bar X}}{\bar \cZ^2} +
\hf \, W^2~.
\label{n=2con}
\ee
Here $\cZ$ denotes the chiral field strength of a vector multiplet which is chosen to be one of the two compensators
of $\cN=2$ Poincar\'e supergravity.\footnote{Within the superconformal tensor calculus, $\cN=2$ Poincar\'e 
supergravity is obtained by coupling conformal supergravity to two compensators, of which one is a vector 
multiplet, and the other can be, e.g.,  a hypermultiplet or a tensor multiplet, see \cite{deWPV} 
and references therein.} The superfield $\cZ$ is reduced chiral, i.e.
it obeys the same equations which $W$ is subject to, 
 (\ref{3}) and \eqref{n=2bi-i}. The superfield $X$ can be expressed in terms of  $W$, $\bar W$ and their 
derivatives by iteratively solving the equation \eqref{n=2con} with $1/\cZ$ considered as a small parameter. 
In the limit $\cZ \to \infty$, the above action reduces to that describing 
Maxwell's action coupled to $\cN=2$ conformal supergravity 
\bea
S_{\rm Maxwell} = \frac{1}{8}  \int \rd^4 x \,{\rm d}^4\q \,\cE\,  W^2
+  \frac{1}{8}  \int \rd^4 x \,{\rm d}^4 \bar \q \, \bar \cE\,  \bar W^2~. 
\eea
The fact that the system defined by eqs. \eqref{21} and \eqref{n=2con} 
is a solution of the self-duality equation \eqref{SDE}, 
can be established  by analogy with the rigid-supersymmetric proof given in \cite{KT1}.
In the rigid supersymmetric limit, the system  \eqref{21} and \eqref{n=2con}  reduces to the one proposed by 
Ketov \cite{Ketov}. The latter is a higher derivative extension of the $\cN=2$ supersymmetric 
Born-Infeld action proposed in \cite{KT2,BIK} as the model for partial 
breaking of supersymmetry $\cN=4 \to \cN=2$ (the model  proposed in  \cite{KT2,BIK} 
is not yet known in a closed form).

Perturbative nonlinear solutions of the
self-duality equation \eqref{SDE} may be constructed similarly to the rigid supersymmetric case 
\cite{KT2,Carrasco:2011jv}. 
We believe our results will be useful in the context of perturbative construction of nonlinear deformations 
of all classically duality invariant theories, including $\cN=8 $ supergravity
(see \cite{Carrasco:2011jv,Broedel:2012gf} and references therein).
\\

\noindent
{\bf Acknowledgements:} The author is grateful to Daniel Butter for reading the manuscript.
This work  is supported in part by the Australian Research Council.

\appendix 

% Equation numbers
\numberwithin{equation}{section}

\section{$\cN = 2$ conformal supergravity} \label{grimmspace}
\setcounter{equation}{0}
\allowdisplaybreaks

This appendix contains a summary of the  superspace formulation for $\cN=2$ conformal supergravity
developed in \cite{KLRT-M1}.  

Conformal supergravity can be realized in a four-dimensional curved $\cN = 2$ superspace
parametrized by local coordinates $z^M = (x^m, \theta^\mu_\imath, \bar{\theta}_{\dot{\mu}}^\imath 
= (\theta_{\mu \imath})^* )$, where $m = 0, 1, ... \ , 3$, $\mu = 1, 2$. $\dot{\mu} = 1, 2$ and $\imath = \1, \2$.
The structure group is chosen to be $\rm SL(2, \dsC) \times SU(2)$, 
and the covariant derivatives $\cD_A = (\cD_a, \cD_\a^i, \bar \cD^\ad_i)$ read
\bea
\cD_A &=& E_A + \Phi_A{}^{kl} J_{kl}+ \hf \Omega_A{}^{bc} M_{bc} \non \\
		  &=& E_A + \Phi_A{}^{kl} J_{kl}+ \Omega_A{}^{\b\g} M_{\b\g} 
		  + \bar{\Omega}_A{}^{ \dot{\b} \dot{\g} } \bar M_{\dot{\b}\dot{\g}}~.
\eea
Here $M_{cd}$ and $J_{kl}$ are the generators of the Lorentz and SU(2) groups respectively, 
and $\O_A{}^{bc}$ and $\Phi_A{}^{kl}$ the corresponding connections. 
The action of the generators on the covariant derivatives are defined as:
\begin{subequations}
\bea
[M_{\a\b},  \cD_\g^i] &=& \ve_{\g (\a } \cD_{\b)}^i \ , \qquad [\bar M_{\ad \bd} , \bar \cD^i_\gd ] 
= \ve_{ \gd (\ad } \bar \cD^i_{\bd) i } ~, \\
\big[ J_{kl}, \cD_\a^i \big] &=& - \d^i_{(k} \cD_{\a l)} \ , \qquad 
[J_{kl}, \bar \cD^\ad_i] = - \ve_{i ( k} \bar \cD^\ad_{l)} \ .
\eea
\end{subequations}

The algebra of covariant derivatives is \cite{KLRT-M1}
\begin{subequations} 
\bea
\{\cD_\a^i,\cD_\b^j\}&=&
4S^{ij}M_{\a\b}
+2\ve^{ij}\ve_{\a\b}Y^{\g\d}M_{\g\d}
+2\ve^{ij}\ve_{\a\b}\bar{W}^{\gd\dd}\bar{M}_{\gd\dd}
\non\\
&&
+2 \ve_{\a\b}\ve^{ij}S^{kl}J_{kl}
+4 Y_{\a\b}J^{ij}~,
\label{acr1} \\
\{\cD_\a^i,\cDB^\bd_j\}&=&
-2\ri\d^i_j(\s^c)_\a{}^\bd\cD_c
+4\d^{i}_{j}G^{\d\bd}M_{\a\d}
+4\d^{i}_{j}G_{\a\gd}\bar{M}^{\gd\bd}
+8 G_\a{}^\bd J^{i}{}_{j}~.
\label{acr2}
\eea
\end{subequations}
Here the real four-vector $G_{\a \ad} $,
the complex symmetric  tensors $S^{ij}=S^{ji}$, $W_{\a\b}=W_{\b\a}$, 
$Y_{\a\b}=Y_{\b\a}$ and their complex conjugates 
$\bar{S}_{ij}:=\overline{S^{ij}}$, $\bar{W}_{\ad\bd}:=\overline{W_{\a\b}}$,
$\bar{Y}_{\ad\bd}:=\overline{Y_{\a\b}}$ obey additional differential constraints implied 
by the Bianchi identities \cite{Grimm,KLRT-M1}.

\footnotesize{

}

\end{document}